\documentclass[12pt]{article}
\pdfoutput=1

\usepackage{color}
\usepackage{amsmath}
\usepackage{amsfonts}
\usepackage{amssymb}
\usepackage{caption}
\usepackage{graphicx}
\usepackage{slashed}            
\usepackage{subfig}             
\usepackage{xspace}				
\usepackage{cite}
\usepackage[english]{babel}
\usepackage{fancyhdr}
\usepackage{amsmath}
\usepackage{amssymb}
\usepackage{amsfonts}
\usepackage{psfrag}
\usepackage[applemac]{inputenc}
\usepackage{graphicx}
\usepackage{braket}
\usepackage{breqn}
\usepackage{feynmp}
\usepackage{feynmp-auto}
\usepackage{subfig}
\usepackage[export]{adjustbox}
\usepackage[toc,page]{appendix}
\usepackage{mathtools} 
\usepackage{float}
\usepackage[
	colorlinks=true,
	citecolor=black,
	linkcolor=black,
	urlcolor=blue,
	hypertexnames=false]{hyperref}

\newcommand{\arXiv}[2]{\href{http://arxiv.org/pdf/hep-ph/#1}{{\tt #2/#1}}}
\newcommand{\arXivold}[2]{\href{http://arxiv.org/pdf/#1}{{\tt #2/#1}}}

\numberwithin{equation}{section} 

\begin{document}
\begin{titlepage}

\begin{center}

	{	
		\LARGE \bf 
		A Phenomenological Approach to Multi-Higgs Production at High Energy
	}
	
\end{center}
	\vskip .3cm

\begin{center} 
{\bf \  Fayez Abu-Ajamieh\footnote{\tt
		 \href{mailto:fayez.abu-ajamieh@umontpellier.fr}{fayez.abu-ajamieh@umontpellier.fr}
\setcounter{footnote}{0}
		 }
		} 
\end{center}

\begin{center} 

	{LUPM UMR5299, Universit\'e de Montpellier, 34095 Montpellier, France}

\end{center}


\centerline{\large\bf Abstract}

\begin{quote}
We tackle the issue of the factorial growth in the amplitudes of multi-Higgs production at high energy by developing a phenomenological approach based on the Higgs splitting functions and Sudakov factors. We utilize the method of generating functionals to define several jet observables for the Higgs sector. Our results suggest that pure Higgs splittings should retain a good Ultraviolet (UV) behavior in contrast to the common picture represented by the breakdown of perturbation theory and the violation of unitarity due to the high multiplicity of particles produced at or near threshold, which is found in scalar theories. We thus argue that the issue of the factorial growth in the amplitude of multi-Higgs production is probably associated with applying perturbation theory in a regime where it is no longer valid and with the $n\lambda \rightarrow \infty$ limit, as opposed to being a sign of new physics.
\end{quote}

\end{titlepage}


\section{Introduction}\label{Chap:Intro}
It has long been known \cite{Cornwall:1990hh} that in a weakly-interacting theory, the production of high-multiplicity final states $n$ at sufficiently high energies leads to the breakdown of perturbation theory when $n \gtrsim 1/\lambda$, where $\lambda$ is the coupling of the theory. This has been studied intensively in theories with scalars  \cite{Goldberg:1990qk, Voloshin:1992mz, Argyres:1992np, Brown:1992ay, Voloshin:1992rr, Voloshin:1992nu, Smith:1992rq, Libanov:1994ug, Gorsky:1993ix, Son:1995wz, Bezrukov:1995ta, Libanov:1997nt}, where it was found that for both the broken and the unbroken phases of $\phi^{4}$ theories, the amplitude of $n$ final state scalars produced at or near threshold through the decay of a highly off-shell initial scalar would grow $\sim n!\lambda^{n}$. This factorial growth leads to an exponential growth in the cross-section after integrating over the phase space:
\begin{equation}\label{eq:Exponentiation}
\sigma_{n} \sim \frac{1}{n!} \int d\Phi_{n} |\mathcal{A}_{1^{*}\rightarrow n}|^{2} \hspace{2mm}\sim n!\lambda^{n} \hspace{2mm}\sim e^{n\log{(\lambda n)}}.
\end{equation}

The factorial growth in the amplitude can be traced to the factorial growth in the number of Feynman diagrams for $\phi^{*} \rightarrow n\phi$, which unlike the case in Quantum Chromodynamics (QCD), lacks destructive interference that would compensate for this factorial growth. It has been argued in the literature that an exponentially growing cross-section would signal the onset of strong dynamics in the weak sector, indicating new physics at high energies. 

Recently, the proposed 100-TeV Future Circular Collider (FCC) has renewed the interest in multi-particle production and in particular in the SM Higgs sector. It was suggested that a very high number of Higgses can be produced near threshold at the scale of tens of TeV, thereby presenting a probe for new physics through the Higgs sector. More specifically, the scattering amplitude of $h^{*} \rightarrow nh$ at threshold in the Higgs sector is given by \cite{Argyres:1992np, Brown:1992ay}
\begin{equation}\label{eq:HiggsAmp}
\mathcal{A}_{1^{*}\rightarrow n} = \Big(\frac{\partial}{\partial z} \Big)^{n} h_{cl} = n!(2v)^{1-n},
\end{equation}
where $h_{cl}$ is the classical solution of the Higgs equation of motion at threshold and $v$ is the Higgs Vacuum Expectation Value (VEV). It was shown in \cite{Libanov:1994ug} that the cross-section would exponentiate in the double-scaling limit
\begin{equation}\label{eq:DoubleScaling}
\sigma_{n} \sim e^{nF(n\lambda,\varepsilon)}, \hspace{0.5 cm} \text{for} \hspace{1 mm} n \rightarrow \infty, \hspace{0.5 cm} n\lambda = \text{fixed}, \hspace{0.5 cm} \varepsilon = \text{fixed},
\end{equation}
where $\varepsilon$ is the average kinetic energy per particle:
\begin{equation}\label{eq:AverageE}
\varepsilon = (E - nM)/(nM),
\end{equation}
and $F(n\lambda,\varepsilon)$ is an approximately known function dubbed "the holy grail" function that includes all contributions to all orders, including loop contributions. It was argued that the exponential cross-section would violate unitarity at high energy (or high multiplicity) thus signaling new physics, (see for example the "Higgsplosion" proposal \cite{Khoze:2015yba, Degrande:2016oan, Khoze:2017tjt, Khoze:2017lft, Khoze:2017ifq} (also see \cite{Khoze:2018mey} for a review)).

The same results were replicated using a semi-classical treatment analogous to instanton-based calculations \cite{Gorsky:1993ix ,Son:1995wz, Khoze:2017ifq}, however, both approaches were derived for the double-scaling limit in Eq. (\ref{eq:DoubleScaling}), which assumes that $n$ would be large ab initio. However, there is no reason for the number of produced Higgses to be large from the beginning, as we show that the probability of producing each extra Higgs should be minuscule, and thus it would be highly unlikely that the Higgs sector will ever enter a non-perturbative regime at colliders.\footnote{In \cite{Gorsky:1993ix}, it was argued that the total probability associated with multi-boson states should rapidly fall with energy in the high-energy regime: $\Gamma (1^{*} \rightarrow n) \sim |\mathcal{A}(1^{*} \rightarrow B)|^{2} \Gamma(B \rightarrow n) \sim e^{-2D(E)}$, where $B$ is an intermediate $N$-state bubble formed by the initial virtual particle, and $D(E)$ is a function of energy which will eventually cut off the amplitude.}

Before proceeding with our approach, we note that in addition to the complication arising from the factorial growth of the final-state Higgs bosons, there is another complication that arises from the production of the intermediate Higgs itself. As discussed in detail in \cite{Jaeckel:2014lya, Khoze:2015yba, Degrande:2016oan}, the production of the Higgs boson is dominated by gluon fusion $gg \rightarrow h^{*} \rightarrow n \times h$, and one needs to include the computation of Feynman diagrams involving 1-loop polygons with $2+k$ edges for all $k \leq n$, where $k$ is the number of outgoing Higgs lines. However, the number of contributing diagrams grows with $n$ and eventually explodes with high multiplicity $n \gg 1$. We stress that this issue is beyond the scope of this work.  Interested readers are referred to \cite{Jaeckel:2014lya, Khoze:2015yba, Degrande:2016oan}.

In this work, we try to approach the issue of multi-Higgs production at high energies differently. We follow a more phenomenological approach to argue that the Higgs sector should retain a good UV behavior at high energy scales relevant to the FCC. Here we try to utilize the success of QCD in describing multi-jet events to the Higgs sector by extending the definition of jets to the Higgs sector. This analogy is motivated by the fact the Higgs quartic coupling $\lambda$ at high energy exhibits a behavior similar to asymptotic freedom in QCD. More specifically, the Renormalization Group Equation (RGE) running of $\lambda$ was calculated up to the Next-to-Next-to-Leading_Order (NNLO) \cite{Chetyrkin:2012rz, Bezrukov:2012sa, Degrassi:2012ry}, and shows that $\lambda$ becomes smaller at higher energies and eventually runs to a fixed point at scales $\gtrsim 10^{9}$ GeV. This behavior is somewhat similar to asymptotic freedom in QCD, in spite of the fact that $\lambda$ does not become non-perturbatively strong in the Infrared (IR) region. In addition, at high energies relevant for the 100 TeV collider, the Higgs can be treated as massless in a manner similar to the case in QCD. This represents enough motivation to extend the QCD treatment to the Higgs sector at high energies. 

To describe our approach more concretely, we imagine an intermediate off-shell Higgs produced with very high energy that subsequently undergoes multiple splittings into several soft Higgses with small transverse momenta. This picture allows us to define a splitting function for the Higgs in a way similar to the QCD splitting functions. If we visualize these radiated soft Higgses (together with their possible decay products) as Higgs "jets", then we can use the splitting functions to resum all the soft splittings radiated off the hard Higgs through the usual Sudakov factor.

The analogy with the QCD sector can be extended to allow for the description of the evolution of the Higgs distribution through the Dokshitzer-Gribov-Lipatov-Altarelli-Parisi (DGLAP) equation \cite{Altarelli:1977zs,Dokshitzer:1977sg,Gribov:1972ri}:
\begin{equation}
\frac{\partial f_{B}(z,\mu^{2})}{\partial \mu^{2}} = \sum_{A} \int_{z}^{1} \frac{d\xi}{\xi}\frac{d\mathcal{P}(z/ \xi,\mu^{2})}{dz dp_{T}^{2}}f_{A}(\xi,\mu^{2}).
\end{equation}

Defining the Higgs distribution through the DGLAP equation allows us to furnish several useful observables that can be used to study the Higgs production at high energies. As we shall see, this picture suggests that the Higgs production should remain well-behaved at high energies, i.e., the number of Higgses produced at high energy should remain low and the Higgs sector should be well-described by the Standard Model (SM). We note here that with this approach, perturbative unitarity is assumed \textit{ab initio} as we will be using perturbation theory implicitly. This is justified as we will do our calculation in the region of the phase space where it remains valid, and use the results as insight to argue that the good behavior should be extrapolated to all regions in the phase space.

We should emphasize here, however, that we are \textit{not} claiming to have solved the factorial divergences problem, which is more of a technical problem associated with Quantum Field Theory (QFT) and perturbation theory. Instead, what we are suggesting is that this problem is probably an artifact resulting from applying perturbation theory (and other semi-classical treatments) in a regime where it breaks down and from assuming the double-scaling limit, and therefore should not be interpreted as a sign of new physics and should not appear in real processes at colliders, at least in the SM Higgs sector.
 
This paper is organized as follows: In Section. \ref{chap:Splitting} we derive the splitting functions of the Higgs cubic and quartic interactions and use them to find the associated Sudakov factors. In Section \ref{chap:Observables} we define a number of Higgs jet observables for both the cubic and the quartic interactions by utilizing the method of generating functionals and show that the average number of Higgses expected at high energy should remain low. We also compare the cubic and the quartic interactions and find that cubic splittings are dominant. We relegate some of the technical details to Appendix \ref{App1}. In Section \ref{Chap:SecondaryEmission} we estimate the contribution of secondary emissions and then we discuss our results and the future outlook in Section \ref{Chap:Conclusions}.

\section{Splitting Functions and the Sudakov Factors}\label{chap:Splitting}

Our starting point will be to derive the splitting functions for the Higgs cubic and quartic interactions and then to use them to find the corresponding Sudakov factors. In doing so, we follow the method originally introduced in \cite{Altarelli:1977zs} and recently utilized by \cite{Chen:2016wkt} to find all of the splitting functions for the entire Electroweak (EW) sector. In all of our calculations, we work in the high energy limit $Q \gg m,v$ (the Higgs mass and VEV), such that all masses can be dropped. However, we do keep the mass as an Infrared (IR) cutoff when we find the Sudakov factors later on. Furthermore, we shall assume the collinear limit where the transverse momentum is small compared with the energy scale of the hard process $p_{T} \ll Q$.

\subsection{The 3-Higgs Vertex}\label{Sec:3H}
\begin{figure}[t!]
\centering
\includegraphics[scale = 0.5]{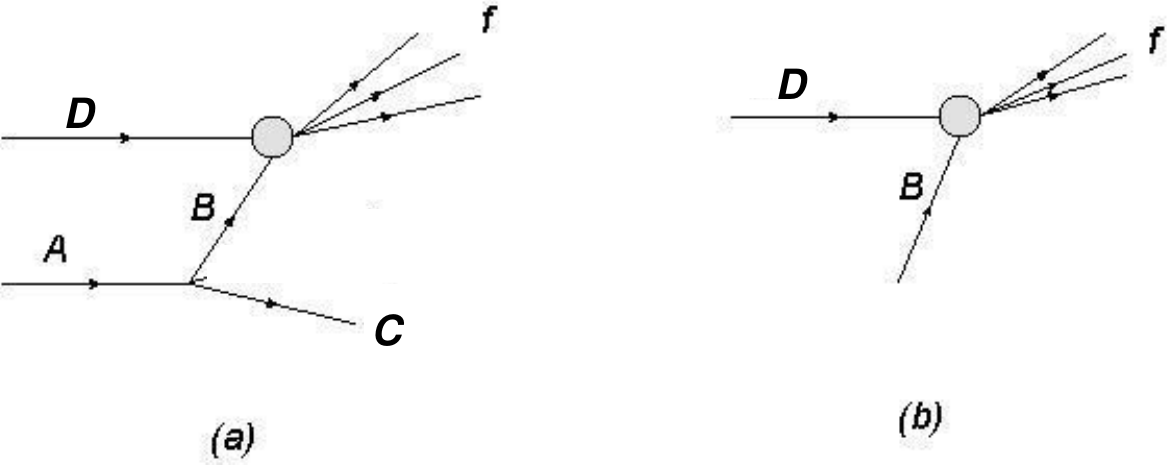}
\caption{Factorization of the 3-vertex splitting.}
\label{fig:3Vertex}
\end{figure}

This section is largely a review of the standard procedure for calculating splitting functions and the Sudakov factor. To derive the splitting function of a general cubic interaction, we consider the processes shown in Fig. \ref{fig:3Vertex}. We assume that the process in (a) is comprised of the hard process in (b) and a soft splitting $A \rightarrow B + C$. Particles $A$ and $B$ are assumed to slightly off-shell with small transverse momenta. Then the differential splitting function  $d\mathcal{P}_{AB}(z)$ is defined as the probability of finding particle $B$ in particle $A$ with an energy fraction $z$ of the energy of $A$ at the lowest order in the coupling:
\begin{equation}\label{eq:3VSplitting0}
d\mathcal{P}_{A\rightarrow BC}(z,p_{T}^{2}) = \frac{\alpha}{2\pi} P_{A\rightarrow BC}(z) dz dp_{T}^{2},
\end{equation}
where $p_{T}$ is the transverse momentum and $P_{AB}(z)$ is the so-called kernel function. The matrix elements of the two processes in Fig. \ref{fig:3Vertex} can be expressed in terms of their interaction vertices as
\begin{subequations}\label{eq:MatrixElements}
\begin{align}
& \mathcal{M}_{A + D \rightarrow C + f}  = g^{2} \frac{V_{A \rightarrow B + C}V_{D+B \rightarrow f}}{(2E_{B})(E_{B}+E_{C}-E_{A})} , \label{eq:3VMatrixElement(a)} \\
& \mathcal{M}_{B + D \rightarrow f}  = g V_{B+D \rightarrow f},\label{eq:3VMatrixElement(b)}
\end{align}
\end{subequations}
where $V_{ij}$ are the invariant matrix elements of the vertices with the factor $(2E_{k})^{-1/2}$ removed and $g$ is the coupling constant. The matrix elements in Eq. (\ref{eq:MatrixElements}) can be used to calculate the cross-sections of the two processes
\begin{subequations}
\begin{align}
& d\sigma_{a} = \frac{g^{4}}{8E_{A}E_{D}} \frac{|V_{A \rightarrow B + C}|^{2}|V_{B+D \rightarrow f}|^{2}}{(2E_{B})^{2}(E_{B}+E_{C}-E_{A})^{2}}(2\pi)^{4} \delta^{4}(K_{A} + K_{D} -K_{C}-K_{f}) \frac{d^{3}\vec{k}_{C}}{(2\pi)^{3}2E_{C}} \prod_{f} \frac{d^{3}\vec{p}_{f}}{(2\pi)^{3}2E_{f}}, \label{eq:3VXsection(a)} \\
& d\sigma_{b} = \frac{g^{2}}{8E_{B}E_{D}} |V_{B+D \rightarrow f}|^{2} (2\pi)^{4} \delta^{4}(K_{B}+K_{D}-K_{f}) \prod_{f} \frac{d^{3} \vec{p}_{f}}{(2\pi)^{3}2E_{f}}. \label{eq:3VXsection(b)}
\end{align}
\end{subequations}
Inspecting eqs. (\ref{eq:3VXsection(a)}) and (\ref{eq:3VXsection(b)}), we can see that they are related in the following way
\begin{equation}\label{eq:3V&4V}
d\sigma_{a} = \frac{E_{B}}{E_{A}} \frac{g^{2}|V_{A \rightarrow B + C}|^{2}}{(2E_{B})^{2}(E_{B}+E_{C}-E_{A})^{2}} \frac{d^{3}\vec{k}_{C}}{(2\pi)^{3}2E_{C}} d\sigma_{b}.
\end{equation}

On the other hand, in the collinear limit where the transverse momenta of particles $B$ and $C$ are small compared to the energy scale of the hard process, the two processes factorize through the differential splitting function \cite{Collins:1989gx}
\begin{equation}\label{eq:3VFactorization}
d\sigma_{a} \simeq d\mathcal{P}_{A \rightarrow B+C}(z,t) \times d\sigma_{b}.
\end{equation}

Comparing eqs. (\ref{eq:3V&4V}) and (\ref{eq:3VFactorization}), we can immediately find a general expression for the splitting function of any cubic interaction:
\begin{equation}\label{eq:3VSplitting1}
d\mathcal{P}_{A \rightarrow BC}(z,t) = \frac{1}{S}\frac{E_{B}}{E_{A}} \frac{g^{2}|V_{A \rightarrow B +C}|^{2}}{(2E_{B})^{2}(E_{B} + E_{C} - E_{A})^{2}} \frac{d^{3}\vec{k}_{C}}{(2\pi)^{3}2 E_{C}},
\end{equation}
where $S$ is a possible symmetry factor. The splitting function depends on a dimensionless variable $z$, which expresses the fraction of the energy of the mother particle that is carried away by the daughter particle (the other daughter particle carries the rest $1-z$), and a dimensionful variable $t$ that expresses the energy scale of the splitting. Common choices of $t$ are the transverse momentum of the daughter particles, the virtuality, or the energy-weighted angle of the radiated particle relative to the mother particle $\theta E_{A}$. In our analysis, we shall use the transverse momentum and set $t \equiv p_{T}^{2}$.

In the collinear limit $|\vec{p}_{T}| \ll Q$, where $Q$ is the energy scale of the mother particle, we can parameterize the 4-momenta of $A$, $B$, and $C$ to the leading order in the transverse momentum as follows:
\begin{subequations}\label{eq:3V4Momenta}
\begin{align}
& K_{A} = \Big( Q,\textbf{0},Q \Big), \\
& K_{B} = \Big( zQ +\frac{p_{T}^{2}}{2zQ},\vec{p}_{T},zQ \Big), \\
& K_{C} = \Big( (1-z)Q +\frac{p_{T}^{2}}{2(1-z)Q},-\vec{p}_{T},(1-z)Q \Big).
\end{align}
\end{subequations}

Notice that particles $B$ and $C$ have virtualities of $O(p_{T}^{4})$. Given this parameterization of momenta, and integrating over the azimuthal angle, we can write the phase space factor as
\begin{equation}\label{eq:3VPhaseSpace}
\frac{d^{3}\vec{k}_{C}}{(2\pi)^{3}2 E_{C}} = \frac{1}{16\pi^{2}}\frac{dz dp_{T}^{2}}{(1-z)}.
\end{equation}

Plugging eqs. (\ref{eq:3V4Momenta}) and (\ref{eq:3VPhaseSpace}) in Eq. (\ref{eq:3VSplitting1}) and keeping only the leading term in $p_{T}^{2}$, the splitting function simplifies to
\begin{equation}\label{eq:3VSplitting2}
\frac{d\mathcal{P}_{A \rightarrow BC}}{dz dp_{T}^{2}} = \frac{1}{S}\frac{g^{2}|V|^{2}}{16\pi^{2}}  \frac{z(1-z)}{p_{T}^{4}}.
\end{equation}

We are now ready to apply this to the Higgs trilinear splitting $h^{*}\rightarrow hh$. Here we work in the normalization $m_{h}^{2} = \frac{1}{2} \lambda v^{2}$, such that $gV_{3H} = \frac{3}{2} \lambda v$. Thus, we finally arrive at the 3H splitting function
\begin{equation}\label{eq:3HSplitting3}
d\mathcal{P}_{h \rightarrow hh}(z,t) = \Big( \frac{3\sqrt{2} v \lambda}{16\pi} \Big)^{2} \frac{z(1-z)}{t^{2}} dz dt.
\end{equation}
\begin{figure}[!t] 
  \centering
  \begin{minipage}[b]{0.48\textwidth}
    \includegraphics[width=\textwidth]{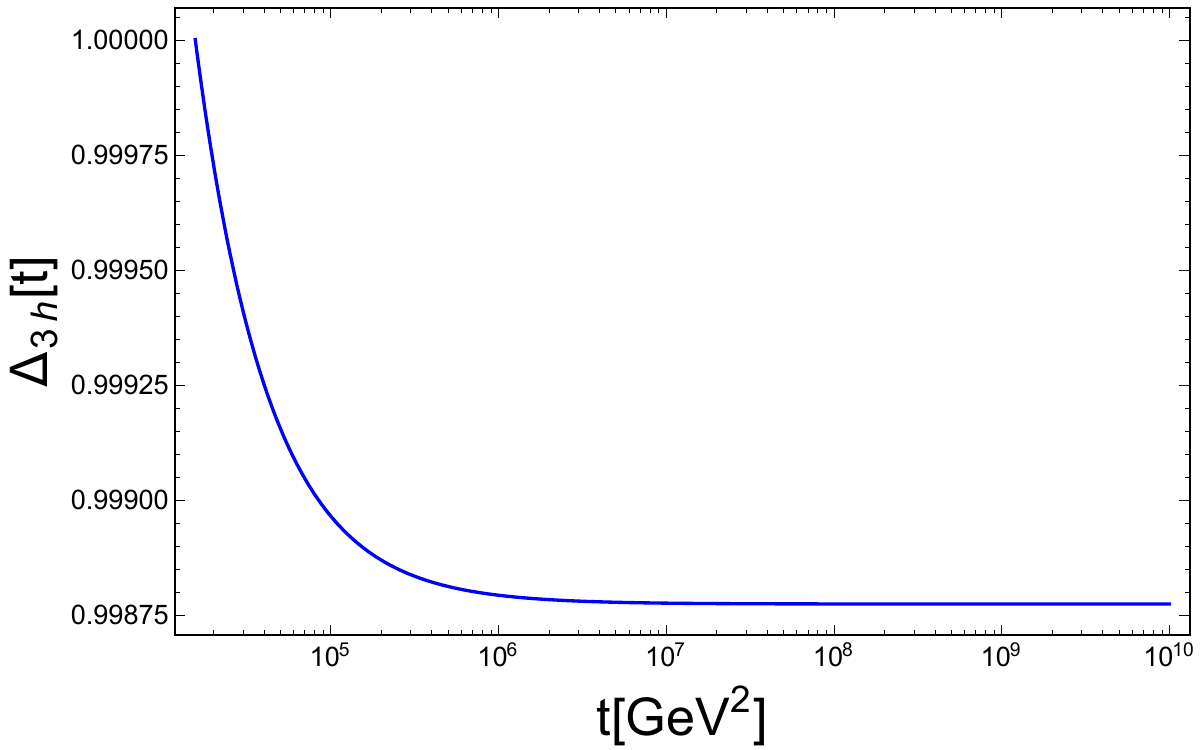}
  \end{minipage}
  \hfill
  \begin{minipage}[b]{0.48\textwidth}
    \includegraphics[width=\textwidth]{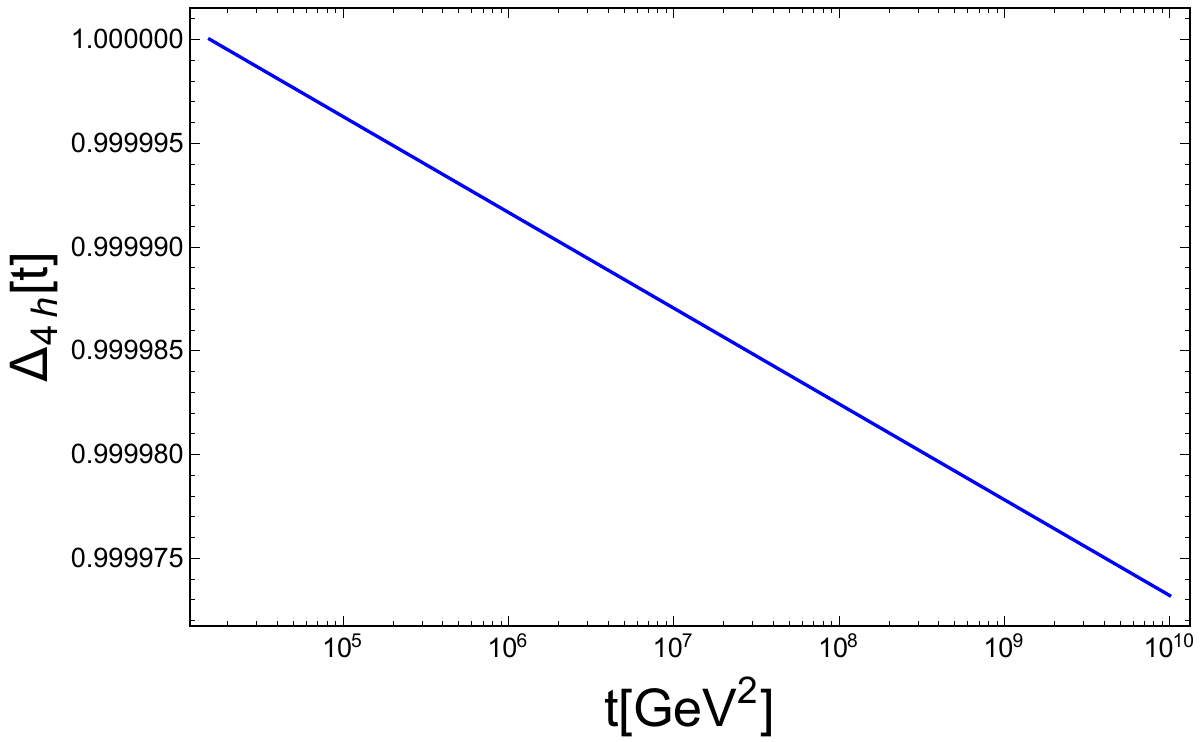}
  \end{minipage}
      \caption{(Left): The cubic Higgs Sudakov factor as a function of the virtuality $t$. (Right): The quartic Higgs Sudakov factor. The plots are on a log-log scale.}
         \label{fig:Sudakovs}
\end{figure}
This result is consistent with \cite{Chen:2016wkt}. Notice here that unlike the splitting functions in the QCD sector, which scales like $\sim dp_{T}^{2}/p_{T}^{2}$, the splitting function of the Higgs cubic interaction scale like $\sim dp_{T}^{2}/p_{T}^{4}$. This type of splitting function dubbed \textit{ultra-collinear} in \cite{Chen:2016wkt} is IR-dominated, with most of the contribution being near $t \sim m^{2}$. Also, integrating these ultra-collinear splitting functions leads to power-law Sudakov factors instead of the usual logarithms as we show below. Finding the Sudakov factor is now a matter of simple integration. Assuming strong-ordering of the radiated particles, the Sudakov factor can be expressed as
\begin{equation}\label{eq:3VSudakov}
\Delta_{3\text{V}} (t,t_{0}) = \text{exp} \Bigg[ - \sum_{BC} \int_{t_{0}}^{t} dt' \int_{0}^{1} dz \frac{d\mathcal{P}_{A \rightarrow B + C}(z,t')}{dz dt'} \Bigg],
\end{equation}
where the sum goes over all particles $B, C$ to which $A$ can decay. Plugging Eq. (\ref{eq:3HSplitting3}) and using the Higgs mass as an IR cutoff, we obtain
\begin{equation}\label{eq:3HSudakov1}
\Delta_{3\text{h}} (t,t_{0}) = \text{exp} \Bigg[ -3\Bigg( \frac{v \lambda}{16\pi} \Bigg)^{2} \Bigg( \frac{1}{t_{0}} - \frac{1}{t}\Bigg) \Bigg],
\end{equation}
where we set $t_{0} = m^{2}$. As noted earlier, the Sudakov factor is dominated near $t \sim m^{2}$ and becomes essentially constant for $t \gg m^{2}$. As the Sudakov factor expresses the probability of a particle \textit{not} splitting, it is easy to see that increasing the energy scale will have a limited effect on enhancing the splitting of the Higgs. This stems from the ultra-collinear behavior of the splitting function which is a direct result of the $dp_{T}^{2}/p_{T}^{4}$ scaling of the splitting function. The Sudakov factor of the trilinear Higgs interaction is shown on the left-hand side of Fig. \ref{fig:Sudakovs}, where we can clearly see that the probability of Higgs splitting remains low even at very high energies. To better understand the smallness of the splitting probability in the Higgs cubic interaction, we write Eq. (\ref{eq:3HSudakov1}) in a more transparent way:
\begin{equation}\label{eq:3HSudakov2}
\Delta_{3\text{h}} (t,t_{0}) = \text{exp} \Bigg[ -\frac{3\alpha_{H}}{8} \Big(1-\frac{m^{2}}{t} \Big) \Bigg],
\end{equation}
where we have define $\alpha_{H} \equiv \lambda/16\pi^{2} \approx 0.003$. We can see that in the limit $t \rightarrow \infty$, $\Delta_{3\text{h}} (t,t_{0}) \rightarrow e^{-3\alpha_{H}/8} \simeq e^{-0.001} \simeq 1$. Thus, we can see that the smallness of the splitting probability is a direct result of the weakness of the Higgs trilinear interaction, coupled with the ultra-collinear behavior of this interaction. This result seems to suggest that one should not anticipate a large number of Higgses in pure Higgs events even at high energies, at least for splittings produced through the trilinear interaction, since the probability of splitting is always small.

\subsection{The 4-Higgs Vertex}\label{Sec:4H}
\begin{figure}[!h]
\centering
\includegraphics[scale = 0.5]{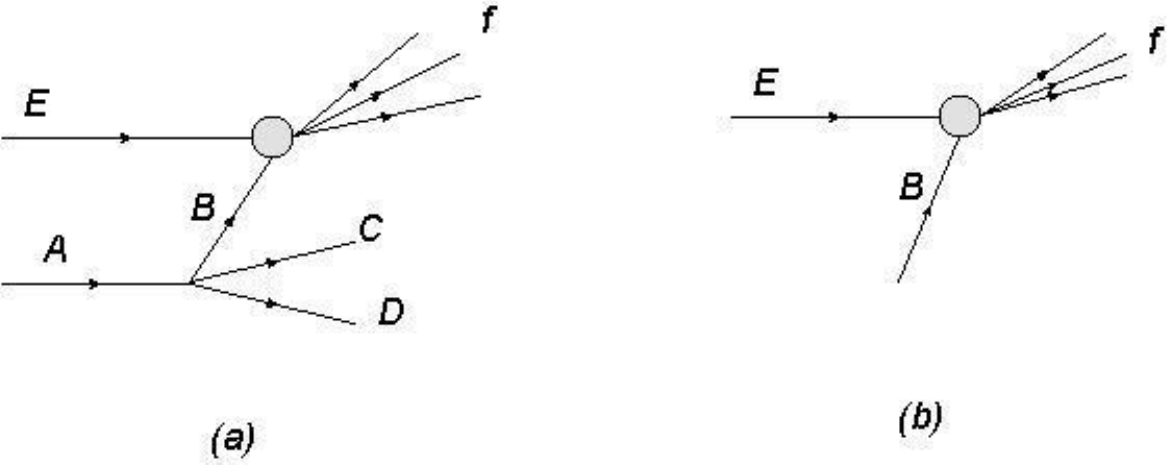}
\caption{Factorization of the 4-vertex splitting.}
\label{fig:4Vertex}
\end{figure}

Now we are in a position to generalize the splitting function and the Sudakov factor to quartic interactions. Previous studies tended to neglect quartic interactions and only focus on cubic terms. We now consider the emission of \textit{two} particles from the same vertex instead of one. Considering the process in Fig. \ref{fig:4Vertex}(a), we can define the quartic splitting function as the probability of finding a pair of particles $C$ and $D$ in particle $A$ with energy fractions $x$ and $z$ of the energy of $A$ at the lowest order of the coupling. The two particles could have different transverse momenta $\vec{p}_{T}$, $\vec{k}_{T}$, and therefore the definition of the splitting function generalizes to:
\begin{equation}\label{eq;4Vsplitting}
d\mathcal{P}_{A \rightarrow BCD}(x,z,p_{T}^{2},k_{T}^{2}) = \frac{\alpha}{2\pi}P_{A \rightarrow BCD} dx dz dp_{T}^{2} dk_{T}^{2}.
\end{equation}

Similarly to the case of the cubic interaction, we assume that the process in Fig. \ref{fig:4Vertex}(a) is comprised of the hard process in (b) and the soft splitting $A \rightarrow B + C + D$.  The matrix elements of the two processes can be written as
\begin{subequations}\label{eq:4VMatrixElements}
\begin{align}
& \mathcal{M}_{E+A \rightarrow C + D +f} = g^{2} \frac{V_{A \rightarrow B + C + D}V_{B + E \rightarrow f}}{(2E_{B})(E_{B}+E_{C}+E_{D}-E_{A})} ,\\
& \mathcal{M}_{E+B \rightarrow f} = g V_{E+B \rightarrow f},
\end{align}
\end{subequations}
and their respective cross-sections are thus given by
\begin{subequations}
\begin{align}
& d\sigma_{a} = \frac{g^{4}}{8E_{A}E_{E}}\frac{|V_{A \rightarrow B + C + D}|^{2}|V_{B+E \rightarrow f}|^{2}}{(2E_{B})^{2}(E_{B}+E_{C}+E_{D}-E_{A})^{2}} \nonumber \\
&  \times (2\pi)^{4} \delta^{(4)}(K_{A} + K_{E} - K_{C} -K_{D} - K_{f}) \frac{d^{3} \vec{k}_{C}}{(2\pi)^{3}2E_{C}} \frac{d^{3} \vec{k}_{D}}{(2\pi)^{3}2E_{D}} \prod_{f} \frac{d^{3}\vec{p}_{f}}{(2\pi)^{3}2E_{f}},  \label{eq:4Vsigma(a)} \\
& d\sigma_{b} = \frac{g^{2}}{8E_{B}E_{E}} |V_{B + E \rightarrow f}|^{2} (2\pi)^{4} \delta^{(4)}(K_{B} + K_{C} - K_{f}) \prod_{f} \frac{d^{3}\vec{p}_{f}}{(2\pi)^{3}2E_{f}}. \label{eq:4Vsigma(b)}
\end{align}
\end{subequations}

Inspecting eqs. (\ref{eq:4Vsigma(a)}) and (\ref{eq:4Vsigma(b)}), and assuming that in the collinear limit $p_{T}, k_{T} \ll Q$, the two process factorize in a way similar to the cubic case in Eq. (\ref{eq:3VFactorization}), and it is not hard to see that the quartic splitting function is given by the following general formula
\begin{equation}\label{eq:4VSplitting0}
d\mathcal{P}_{A \rightarrow BCD}(x,z,p_{T}^{2},k_{T}^{2}) = \frac{1}{S}\frac{E_{B}}{E_{A}} \frac{g^{2}|V_{A \rightarrow B + C + D}|^{2}}{(2E_{B})^{2}(E_{B}+E_{C}+E_{D}-E_{A})^{2}} \frac{d^{3} \vec{k}_{C}}{(2\pi)^{3}2E_{C}} \frac{d^{3} \vec{k}_{D}}{(2\pi)^{3}2E_{D}}.
\end{equation}

This equation is similar to the cubic case, except now it has two energy fractions $x$ and $z$ (with $x+z =1$) and two energy scales $\vec{p}_{T}$, $\vec{k}_{T}$. The 4-momenta of the particles can be parameterized as
\begin{subequations}\label{eq:4V4Momenta}
\begin{align}
& K_{A} = \Big( Q,\textbf{0},Q \Big), \\
& K_{D} = \Big( x Q + \frac{p_{T}^{2}}{2 x Q}, \vec{p}_{T},x Q \Big), \\
& K_{C} = \Big( z Q + \frac{k_{T}^{2}}{2 z Q}, \vec{k}_{T},z Q \Big), \\
& K_{B} = \Big( (1-x-z) Q + \frac{(\vec{p}_{T}+\vec{k}_{T})^{2}}{2(1-x- z) Q}, -\vec{p}_{T} -\vec{k}_{T},(1-x-z) Q \Big).
\end{align}
\end{subequations}

Notice that $\vec{p}_{T}$ and $\vec{k}_{T}$ could have different directions and that the azimuthal angle $\phi$ between them needn't be small even in the collinear limit. In fact, $\phi$ could have any value between $0$ and $2\pi$. This is because the orientations of the emitted particles are independent of the angles $\theta_{i}$ between their individual directions and that of the mother particle $A$, which \textit{are} small in the collinear limit. Thus, the azimuthal dependence can be integrated in one of the phase space factors, but not in both
\begin{subequations}\label{eq:4VPhaseSpace}
\begin{equation}
\frac{d^{3}\vec{k}_{C}}{(2\pi)^{3}2E_{C}} = \frac{dx dp_{T}^{2}}{16\pi^{2} x},
\end{equation}
\begin{equation}
\frac{d^{3}\vec{k}_{D}}{(2\pi)^{3}2E_{D}} = \frac{dz d\phi dk_{T}^{2}}{32\pi^{3} z}.
\end{equation}
\end{subequations}

Putting all pieces together, and keeping only the leading terms in the transverse momenta, Eq. (\ref{eq:4VSplitting0}) simplifies to the following general formula
\begin{equation}\label{eq:4VSplitting1}
\frac{d\mathcal{P}_{A \rightarrow BCD}}{dx dz d\phi dp_{T}^{2}d k_{T}^{2}} = \frac{1}{2\pi S} \Big( \frac{g|V|}{16\pi^{2}} \Big)^{2} \frac{ x z (1-x-z)}{[z(1-z)p_{T}^{2}+x(1-x)k_{T}^{2} + 2 x\hspace{0.5mm} z\hspace{0.5mm} p_{T}\hspace{0.5mm} k_{T} \cos{\phi}]^{2}} \hspace{1mm},
\end{equation}
where $S$ is a possible symmetry factor. To apply this to the Higgs quartic interaction, we insert $g|V_{4H}| = \frac{3}{2} \lambda$ (in the normalization adopted above) and set $S=3$, we obtain
\begin{equation}\label{eq:4VSplitting2}
\frac{d\mathcal{P}_{h \rightarrow hhh}}{dx dz d\phi dp_{T}^{2} dk_{T}^{2}} = \Big( \frac{\sqrt{3} \lambda}{32\sqrt{2\pi} \pi^{2}} \Big)^{2}  \frac{ x z (1-x-z)}{[z(1-z)p_{T}^{2}+x(1-x)k_{T}^{2} + 2 x\hspace{0.5mm} z\hspace{0.5mm} p_{T}\hspace{0.5mm} k_{T} \cos{\phi}]^{2}} \hspace{1mm}.
\end{equation}

Before we use the splitting function to find the Sudakov factor, there is a subtlety that we need to address: In cubic splittings, there is a single well-defined energy scale $p_{T}^{2} \equiv t$, however, for quartic splitting we have two energy scales $p_{T}^{2}, (k_{T}^{2}) \equiv t', (t'')$. Therefore, we first need to generalize Eq. (\ref{eq:3VSudakov}) to the case of quartic interactions. We write
\begin{equation} \label{eq:4VSudakov1}
\Delta_{4V}(t_{0},t)  = \text{exp} \Bigg[ - \sum_{BCD} \int_{0}^{1}dz \int_{0}^{1-z} dx \int_{0}^{2\pi} d\phi \int_{t_{0}}^{t} dt' \int_{t_{0}}^{t'} dt'' \frac{d\mathcal{P}}{dx dz d\phi dt' dt''} \Bigg],
\end{equation}
where the sum should go over all quartic splittings that the mother particle $A$ could undergo. Now we are in a position to use Eq. (\ref{eq:4VSplitting2}) to find the Sudakov factor for the Higgs quartic interaction. The integrals over the energy scales can be done exactly giving the familiar logarithmic factor, while the remaining integrals contain a complicated function of the energy fractions and the azimuthal angle and can be done numerically. The final quartic Higgs Sudakov factor reads
\begin{equation}\label{eq:4HSudakov}
\Delta_{4H}(t,t_{0})  = \text{exp} \Bigg[ - \frac{3b}{8\pi} \alpha_{H}^{2} \log(t/t_{0}) \Bigg],
\end{equation}
where the numerical factor $b \simeq 1.57$ comes from integrating over $x, z$ and $\phi$. Comparing the Sudakov factor of the Higgs cubic splitting with that of the quartic splitting, a couple of remarks are in order: (1) The Higgs quartic splitting exhibits the usual logarithmic scaling instead of the power-law scaling that we found in the cubic Higgs case. This logarithmic scaling is a result of the additional integral over the extra energy scale, and (2) the Sudakov factor of the quartic interaction contains an extra phase space factor of $1/16\pi^{2}$ which leads to a significant suppression relative to the cubic Sudakov factor. We plot the Higgs quartic Sudakov factor on the right-hand-side of Fig. \ref{fig:Sudakovs} where we can see that relative to the cubic Higgs case, the probability of quartic splittings is much smaller due to the extra phase space factor. We will discuss this suppression in more detail in the next section.

\section{Higgs Generating Functionals and Jet Observables}\label{chap:Observables}
Having defined the splitting functions and Sudakov factors for the Higgs cubic and quartic interactions, we would like to treat the Higgses as jets and define several IR-safe jet observables that can be used to investigate the production of multi-Higgses at high energy. To this end, we shall apply the method of generating functionals used for studying QCD jets \cite{Dokshitzer:1991wu, Ellis:1991qj} to the Higgs sector. The method of the generating functionals simply aims at constructing an $n$-particle functional in an arbitrary parameter $u$, whose repeated differentiation with respect to $u$ yields the cross-sections of the $n$-particles as the coefficients of the expansion. Thus, the generating functional can be constructed by summing all tree-level cross-sections weighted by an appropriate power of $u$. In the following, we follow a construction more suitable for our purposes presented in \cite{Gerwick:2012hq} (see also \cite{Plehn:2009nd}). When we divide the contributions by the total cross-section, then the repeated differentiation yields the exclusive multiplicity distribution $P_{n} = \frac{\sigma_{n}}{\sigma_{tot}}$. Thus, the generating functional is constructed as follows:
\begin{equation}\label{eq:GeneratingFunctional}
\Phi = \sum_{n=1}^{\infty}u^{n}P_{n-1} \hspace{1 cm} \text{where} \hspace{1 cm} P_{n-1} = \frac{\sigma_{n-1}}{\sigma_{\text{tot}}} = \frac{1}{n!}\frac{d^{n}}{du^{n}}\Phi \Bigg|_{u = 0}.
\end{equation}

Note here that $P_{n-1}$ describes $n-1$ radiated jets, i.e. $n=1$ corresponds to the original particle not splitting. We can see that $P_{n-1}$ expresses the relative contribution of each additional radiated particle to the total cross-section. Another important observable that can be extracted from the generating functional that is relevant for our purposes is the average jet multiplicity, which describes the average number of radiated particles at a given energy scale
\begin{equation}\label{eq:AJM1}
\bar{n} = \frac{d\Phi}{du}\Bigg|_{u =1} = \sum_{n=1}^{\infty} n u^{n-1} \frac{\sigma_{n-1}}{\sigma_{\text{tot}}} \Bigg|_{u=1} = 1 + \frac{1}{\sigma_{\text{tot}}} \sum_{n=1}^{\infty}(n-1)\sigma_{n-1}.
\end{equation}

The generating functional method can also be used to study the jet scaling pattern, which simply expresses the relative suppression associated with each additional radiated particle. The jet scaling pattern can be expressed as the ratios of the successive exclusive jet cross-sections
\begin{equation}\label{eq:ScalingPatern1}
R_{(n+1)/n} \equiv \frac{\sigma_{n+1}}{\sigma_{n}} = \frac{P_{n+1}}{P_{n}}.
\end{equation}

The scaling pattern was investigated for the case of QCD jets in \cite{Gerwick:2012hq, Gerwick:2011tm}. In QCD jets, there are two main limiting cases that describe the jet scaling pattern. If the ratio of the successive cross-sections is constant, then the pattern is referred to as a staircase pattern. On the other hand, the pattern is called Poisson if it follows a Poisson distribution:
\begin{equation}\label{eq:Poisson}
P_{n} = \frac{\bar{n}^{n} e^{-\bar{n}}}{n!} \implies R_{(n+1)/n}  = \frac{\bar{n}}{n+1}.
\end{equation}

Below, we derive these observables for the Higgs cubic and quartic interactions and use them to investigate the production of multi-Higgses at high energies.

\subsection{The Higgs Cubic Interaction}
To derive the generating functional, we will follow the method presented in \cite{Plehn:2009nd}. The DGLAP equation describes the evolution of parton densities in QCD. Thus, they can be used to describe parton splittings $i \rightarrow jk$ where each jet is described by the generating functional instead of the parton density. We can thus write the general formula describing the evolution of the generating functionals as
\begin{equation}\label{eq:GFEvolation}
\Phi_{i}(t) = \Delta_{i}(t,t_{0}) \Phi_{i}(t_{0}) + \int_{t_{0}}^{t}dt' \Delta_{i}(t,t')\sum_{i\rightarrow jk} \int_{0}^{1} dz \frac{d\mathcal{P}}{dzdt'}\Phi_{j}(z^{2}t') \Phi_{k}((1-z)^{2}t').
\end{equation}

Given the splitting function and the Sudakov factor that describe a certain splitting, the generalization to any sector will be straightforward. Using the results found earlier, we find the generating functional of the cubic Higgs interaction
\begin{equation}\label{eq:3HGF}
\Phi_{3h}(t) = u \Big[ \Delta_{3h}(t,t_{0}) \Big]^{1-u}.
\end{equation}

The detailed derivation is presented in Appendix \ref{App1}. Eq. (\ref{eq:3HGF}) can be used in eqs (\ref{eq:GeneratingFunctional}), (\ref{eq:AJM1}) and (\ref{eq:ScalingPatern1}) to find the exclusive multiplicity distribution, average jet multiplicity and jet scaling pattern respectively
\begin{subequations}
\begin{align}
& P_{n-1} =\Delta_{3h}(t,t_{0}) \frac{|\log{\Delta_{3h}(t,t_{0})}|^{n-1}}{(n-1)!}, \label{eq:3hEMD} \\
& \bar{n} = 1 - \log{\Delta_{3h}(t,t_{0})}, \label{eq:3hAJM} \\
& R_{(n+1)/n} = \frac{|\log{\Delta_{3h}(t,t_{0})}|}{n+1}. \label{eq:3hScalingPatter}
\end{align}
\end{subequations}

Before we study these observables, we point out a few remarks: (1) Since $\Delta_{3h}(t,t_{0}) \leq 1$, we can see from Eq. (\ref{eq:3hAJM}) that $\bar{n} \geq 1$,\footnote{Notice that in Eq. (\ref{eq:Poisson}), $\bar{n}$ refers to the average number of \textit{radiated} particles, while in Eq. (\ref{eq:3hAJM}) it refers to the \textit{total} number of jets, including the original one.} with the average jet multiplicity being equal to unity only when $t=t_{0}$. This simply means that $t = t_{0}$ corresponds to the original Higgs not splitting, while the number of radiated Higgses is enhanced with increasing the energy scale, and (2) from Eq. (\ref{eq:3hScalingPatter}), we can see that the cubic Higgs splitting follows a Poisson pattern.
\begin{figure}[!t] 
  \centering
  \begin{minipage}[b]{0.48\textwidth}
    \includegraphics[width=\textwidth]{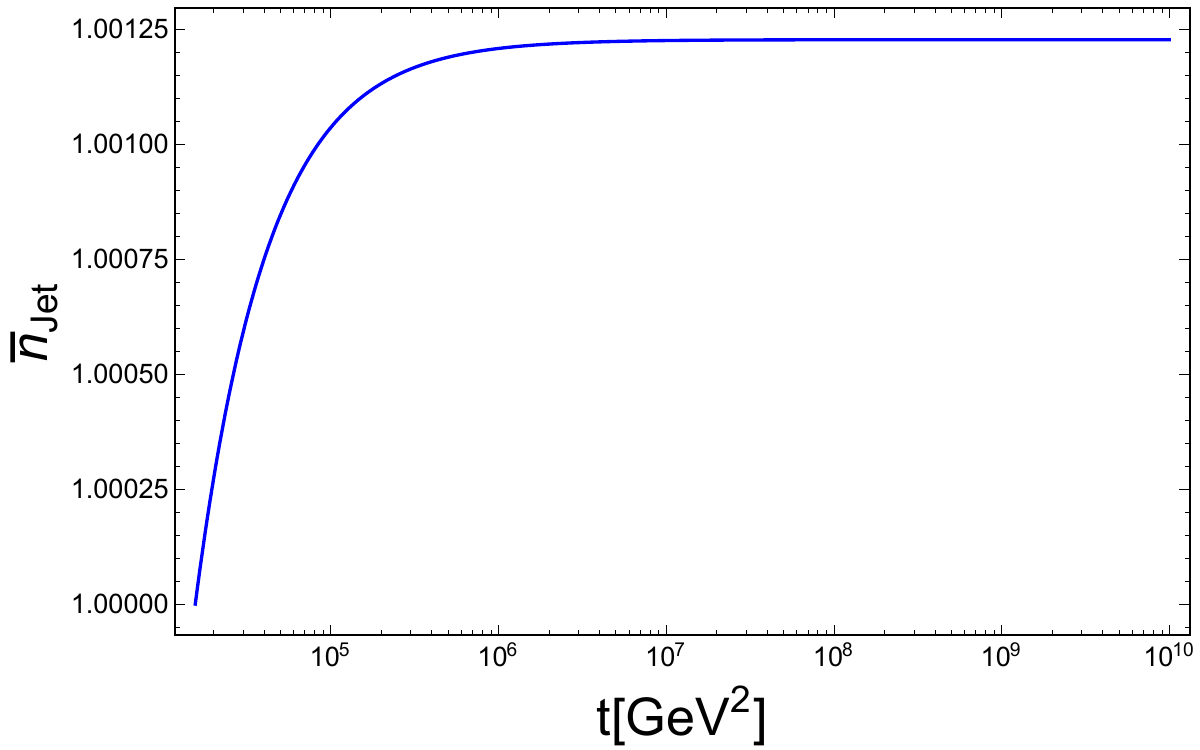}
  \end{minipage}
  \hfill
  \begin{minipage}[b]{0.48\textwidth}
    \includegraphics[width=\textwidth]{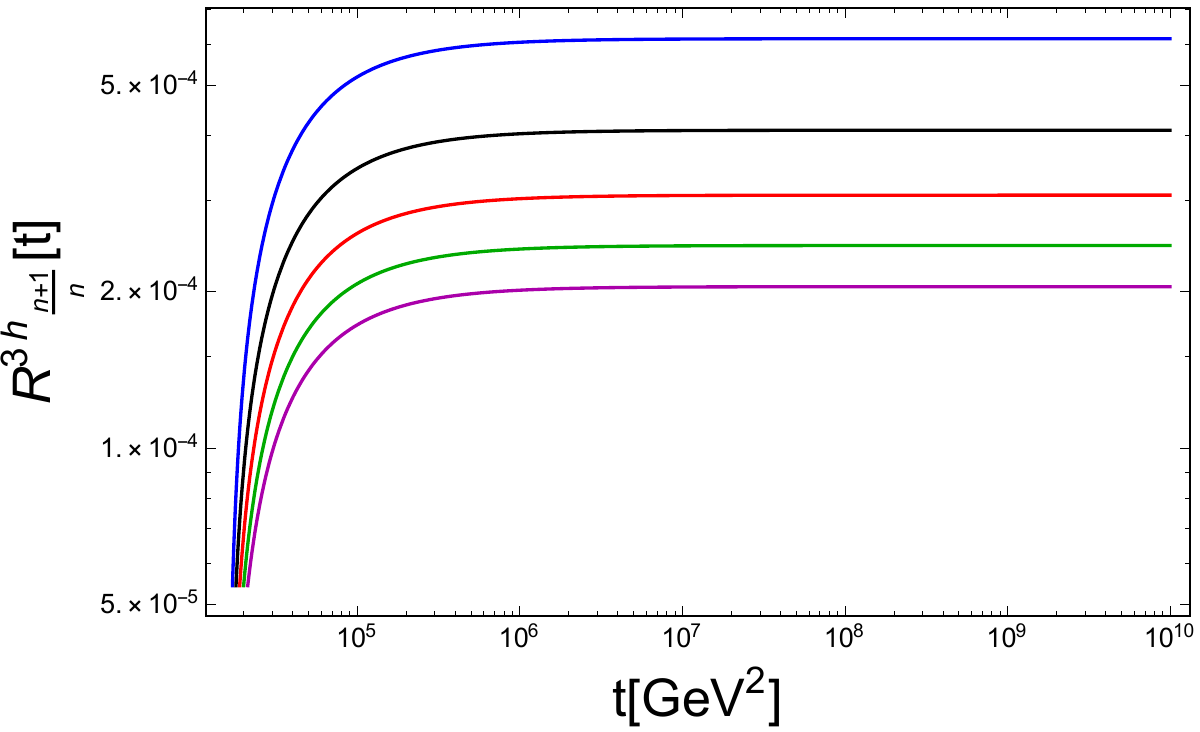}
  \end{minipage}
      \caption{(Left): The cubic Higgs average jet multiplicity with the energy scale. (Right): The cubic Higgs jet scaling pattern given in Eq. (\ref{eq:3hScalingPatter}), with $n=1$ (blue), $n=2$ (gray), $n=3$ (red), $n=4$ (green), $n=5$ (magenta). The plot clearly shows a Poisson scaling pattern (see Eq. (\ref{eq:Poisson})). Both plots are on a log-log scale.}
         \label{fig:3Hobservables}
\end{figure}

The cubic Higgs average jet multiplicity is shown on the left side of Fig. \ref{fig:3Hobservables}. The plot clearly shows that even at very high energy scales, the average number of Higgses is very close to one, i.e. the average number of radiated Higgses is always small, and that most Higgs events will not undergo any splitting (at least through the trilinear interaction). This picture is in stark contrast with the conclusion that a high multiplicity of Higgses would be produced at high energies due to the factorial growth in the amplitude, as highlighted in the introduction. We are thus led to believe that the Higgs sector should remain well-behaved at high energies, and that concluding that new physics should emerge in the Higgs sector at high energy as a result of the supposed factorial growth of the amplitude is probably the wrong conclusion to draw. To put this in more concrete terms, we argue that at high energies, the multi-Higgs production in pure Higgs events should remain perturbative and well-described by the SM; and that the factorial growth in the amplitudes of multiple Higgses produced at or near threshold is probably an artifact of applying perturbation theory where it is not valid, and of assuming the double-scaling limit in Eq. (\ref{eq:DoubleScaling}). Thus, it should not be interpreted as a sign of new physics and should not appear in real processes in colliders.

We should point out, however, that our results are approximate as we are only resumming a subset of the possible $n!$ Feynman diagrams through the Sudakov factor. Therefore, one might argue that other topologies might drastically enhance the Higgs production. For instance, it was argued in \cite{Khoze:2017lft} that the leading contribution to the amplitude stems from the interference terms among the different Feynman diagrams, which contribute an additional $n!$ to the amplitude, however, the exponential growth in the amplitude implicitly assumes that the number of particles that are produced is already large, which is probably not the case. In spite of our approximate treatment, we should emphasize that the differential probability of splitting, as represented by the splitting function, is independent of the topology of the Feynman diagram, and since the probability of splitting is always small, other topologies should not exhibit drastically different behavior. Another approximation in our calculation is the assumption of the collinear limit, which could impact our results. Nonetheless, this assumption is quite justified in the high energy limit wherein we are interested. Therefore, we conclude that the Higgs sector should remain under control at high energy.

We should also point out that we are working in the leading order of $\lambda$ and we are neglecting its RGE running, however, as $\lambda$ becomes smaller at higher energies (see for instance Figure 1 in \cite{Degrassi:2012ry}), then the probability of splitting will become even lower, thereby making the average jet multiplicity even lower than what we find in our calculation with the running neglected. This gives further reasons to believe that the number of Higgses produced at high energies should not become large.

We must, however, emphasize that our results do not represent a solution to the technical problem of the factorial growth in scalar amplitudes in the high multiplicity limit. What we argue here is that this behavior (at least for the Higgs sector), is not a sign of new physics, but rather a limitation of perturbation theory itself and of assuming the double-scaling limit and that for all practical purposes we should trust the predictions of the SM at high energies (at least energies relevant for colliders). Our results are in line with the argument recently presented in \cite{Dine:2020ybn}, where they presented an entirely different, semi-classical non-perturbative treatment for the production of a large number of scalars in the processes $2 \rightarrow n$ and $n \rightarrow n$ in a non-broken $\phi^{4}$ theory. Their results also suggest that using perturbation theory in the regime $n \gtrsim \lambda^{-1}$ is erroneous and that the growth in amplitude is weaker than $n!$. Furthermore, our results are also reminiscent of the results in \cite{Belyaev:2018mtd}, where it is argued that: 1) The formula for Higgsplosion has limited applicability and that it is inconsistent with the unitarity of the SM, and 2) it is not possible to resum the contribution from Higgsplosion in the imaginary part of the Higgs boson propagator, therefore a solution to the hierarchy problem cannot be furnished with this mechanism. We will show below that including the Higgs quartic interaction will not alter this conclusion.

To conclude this subsection, we show the jet scaling pattern for the cubic Higgs interaction on the right side of Fig. \ref{fig:3Hobservables}, where we see that the Poisson pattern is manifest. 

\subsection{The Higgs Quartic Interaction}
Here we perform the same analysis for the quartic Higgs sector. The generalization of the DGLAP equation for generating functionals to the quartic Higgs interaction is fairly straightforward, and the calculation of the generating functional follows the same logic as that for the 3H case. The 4H generating functional is given by
\begin{equation}\label{eq:4hGF}
\Phi_{4h}(t) = u \hspace{1 mm} \Big[ \Delta_{4h}(t,t_{0}) \Big]^{1-u^{2}}.
\end{equation}
\begin{figure}[!t] 
  \centering
  \begin{minipage}[b]{0.48\textwidth}
    \includegraphics[width=\textwidth]{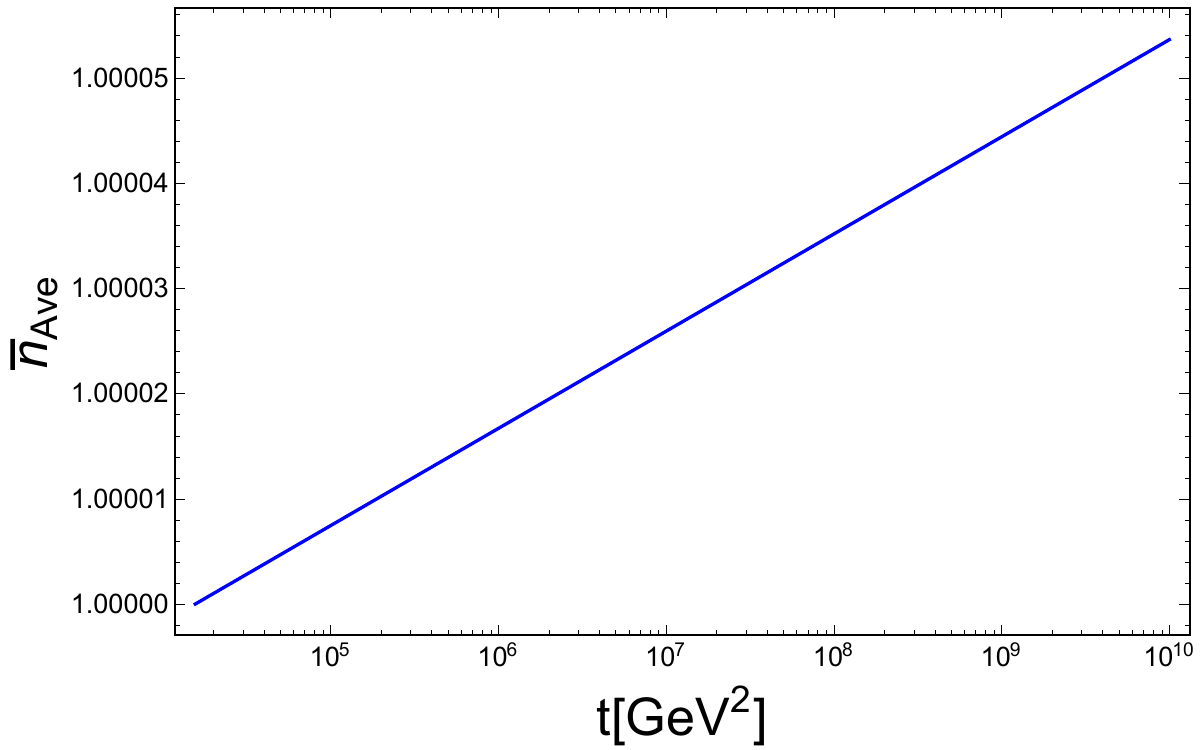}
  \end{minipage}
  \hfill
  \begin{minipage}[b]{0.48\textwidth}
    \includegraphics[width=\textwidth]{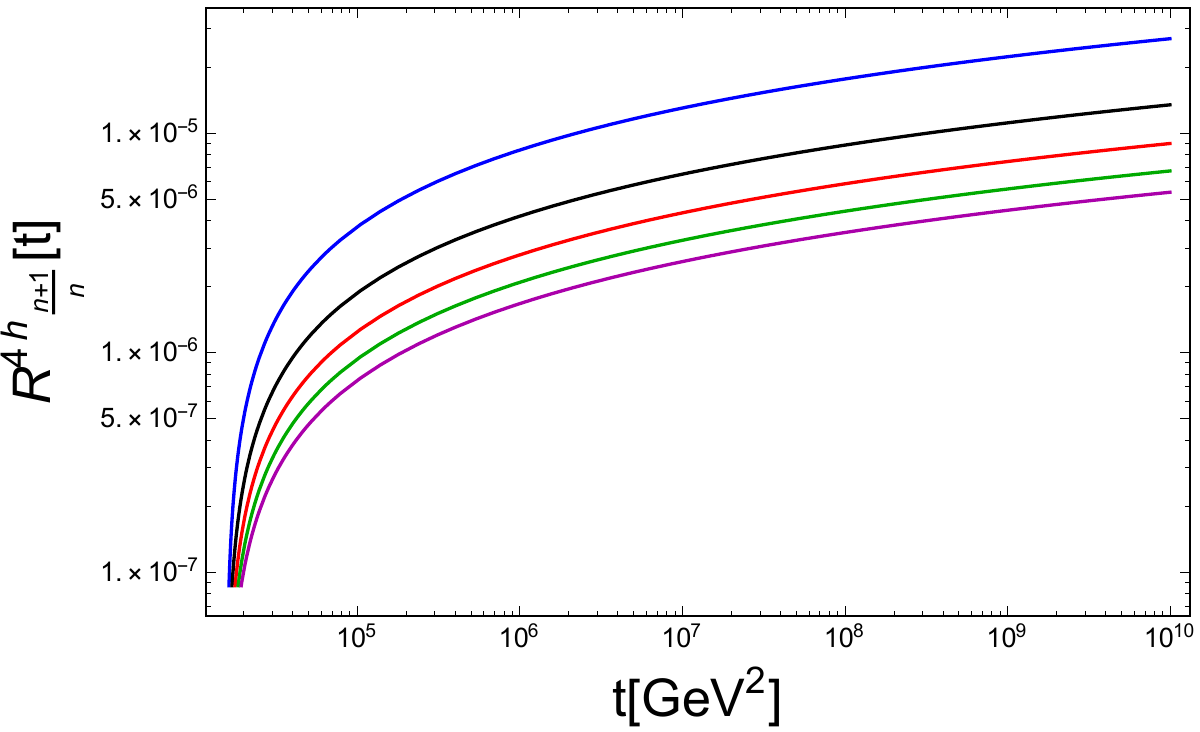}
  \end{minipage}
      \caption{(Left): The quartic Higgs average jet multiplicity with the energy scale. (Right): The quartic Higgs jet scaling pattern given in Eq. (\ref{eq:4hScalingPattern}),  with $n=1$ (blue), $n=2$ (gray), $n=3$ (red), $n=4$ (green), $n=5$ (magenta). The plot clearly shows a Poisson scaling pattern (see Eq. (\ref{eq:Poisson})). Both plots are on a log-log scale}
         \label{fig:4hObservables}
\end{figure}

The 4H generating functional is very similar to the 3H one, with the only difference being in the power of $1-u^{2}$ instead of $1-u$. This is because, in a quartic splitting, \text{two} particles are radiated from the same vertex instead of one. The jet observables can be easily found
\begin{subequations}
\begin{align}
&  P_{n-2} =
  \begin{cases}
                                   \frac{n!!}{n!} \Delta_{4h}(t,t_{0}) \Big[ -2 \log{\Delta_{4h}(t,t_{0})} \Big]^{\frac{n-1}{2}} & \text{;   $n= \text{odd}$}, \\
                                   0 & \text{;   $n=\text{even}$}, \\
 \end{cases} \label{eq:4hEMD}\\
& \bar{n} \hspace{0.1 cm} = \hspace{0.1 cm}  1 - 2\log{\Delta_{4h}(t,t_{0})}, \label{eq:AJM} \\
& R_{(n+2)/n} \hspace{0.1 cm}  = \hspace{0.1 cm} \frac{P_{n+2}}{P_{n}}  \hspace{0.1 cm}  = \hspace{0.1 cm} \frac{|2\log{\Delta_{4h}(t,t_{0})}|}{n+1} \hspace{1.6 cm}  ;n = \text{odd}, \label{eq:4hScalingPattern}
\end{align}
\end{subequations}
and here we see that the jet observables are only defined for an odd number of jets corresponding to an even number of radiated Higgses (2 per splitting) in addition to the original hard Higgs. Here too we find that $\bar{n}_{\text{jet}}\geq 1$ and that the scaling pattern is of Poisson type.

We plot the average jet multiplicity and the jet scaling pattern for the quartic Higgs interaction in Fig. \ref{fig:4hObservables}. Here too we see that the average number of radiated Higgses is minuscule, thereby confirming our earlier conclusion of a good UV behavior of pure Higgs events. Comparing the average jet multiplicities through the cubic and quartic interactions, we find that the cubic interaction dominates. This is hardly surprising as the quartic splitting function has an extra phase space factor of $1/16\pi^{2}$ that exponentiates in the Sudakov factor, thus providing significant suppression, as mentioned in the previous section.

To compare the average jet multiplicities more rigorously, we recall, that the number of splittings $n_{s} = \bar{n}_{\text{jet}}-1$. Thus, we can define the splitting fraction for a certain vertex as
\begin{equation}\label{eq:SplittingFraction}
\beta_{h_{i}} = \frac{n_{s_{i}}}{n_{s_{3h}} + n_{s_{4h}}}.
\end{equation}
\begin{figure}[!t] 
  \centering
    \includegraphics[width=0.6\textwidth]{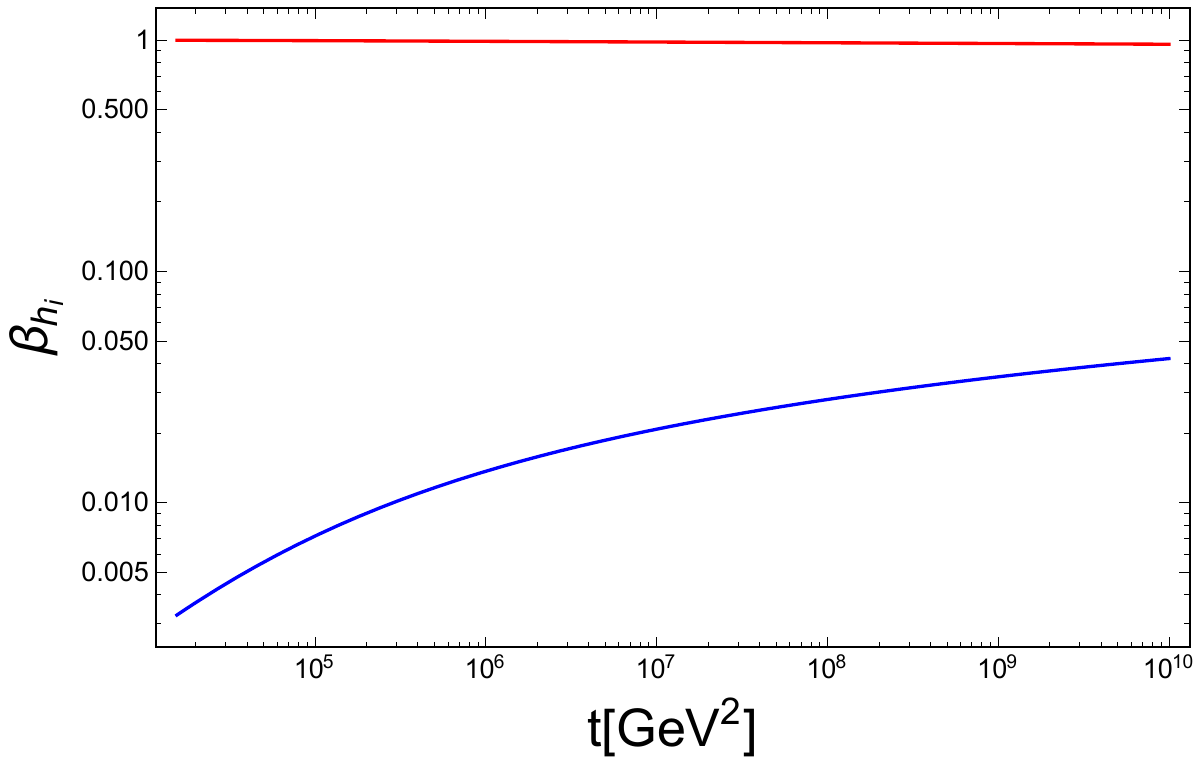}
      \caption{Splitting fractions of the Higgs cubic (red) and quartic (blue) interactions. Notice that $\beta_{3h} + \beta_{4h} =1$. The plot shows that for moderate energies the cubic splitting dominates over the quartic one, while the quartic only begins to dominate at energies $\gtrsim 8 \times 10^{134}$ GeV (not shown on the plot), see Eq. (\ref{eq:UVscale}). The plot is on a log-log scale.}
         \label{fig:SplittingFractions}
\end{figure}

We plot the splitting fraction in Fig. \ref{fig:SplittingFractions}. The plot shows that the cubic Higgs splitting dominates over the quartic one. However, we can also see that the relative contribution of the quartic splitting grows with energy. To estimate the energy scale at which the quartic splitting begins to dominate, we can compare the quartic Sudakov factor (Eq. (\ref{eq:4HSudakov})) with the cubic one (Eq. (\ref{eq:3HSudakov2})). For $t_{0} = m^{2}$, one finds that the quartic scale begins to dominate at an energy scale of:
\begin{equation}\label{eq:UVscale}
Q \simeq	m \hspace{1mm} \text{exp}\Big( \frac{4\pi^{3} v^{2}}{b \hspace{1mm} m^{2}}\Big) =  m \hspace{1mm} \text{exp} \Big( \frac{\pi}{2 \hspace{1mm}b \hspace{1mm} \alpha_{H}}\Big)\simeq 8 \times 10^{134} \hspace{1mm}\text{GeV!}
\end{equation}
thus, for all practical purposes, we can completely neglect the Higgs quartic splittings.

\section{Primary vs. Secondary Emissions}\label{Chap:SecondaryEmission}
\begin{figure}[!t] 
  \centering
    \includegraphics[width=0.7\textwidth]{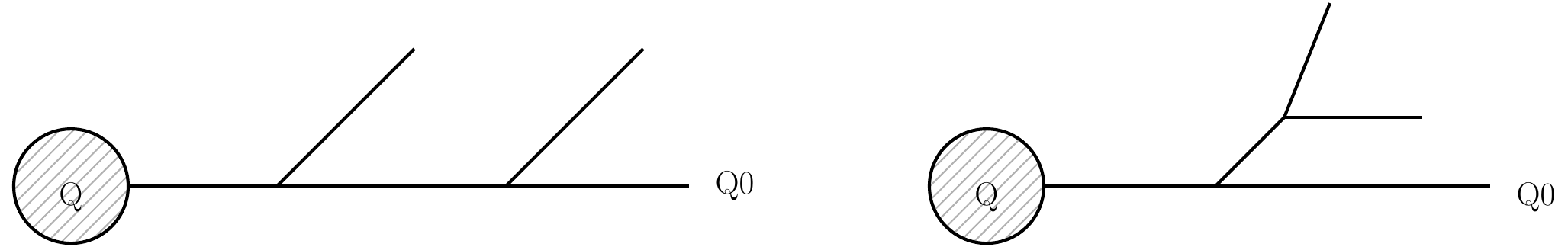}
      \caption{Primary emission (left) vs. secondary emission (right).}
       \label{fig:PrimaryVSecondary}
\end{figure}

So far, we have only considered primary emissions and neglected secondary ones. What we mean by primary emissions are the emissions characterized by the hard Higgs radiating successive soft Higgses. On the other hand, secondary emissions refer to the ones where the soft Higgses themselves radiate other soft Higgses (see Fig. \ref{fig:PrimaryVSecondary}). For the case of QCD jets, primary emissions dominate at high energy, while at low energy it is the secondary emissions that dominate \cite{Gerwick:2012hq}. In the Higgs sector, we would like to estimate how much uncertainty is associated with neglecting secondary emissions. To estimate the contribution of primary and secondary emissions in pure Higgs splittings, we can calculate their cross-sections as follows:
\begin{subequations}
\begin{equation}\label{eq:primary}
\sigma^{\text{P}}(Q^{2},Q_{0}^{2}) = C^{\text{P}} \int_{Q_{0}^{2}}^{Q^{2}} dt \Gamma(Q^{2},t) \Delta_{h}(t,t_{0}) \int_{Q_{0}^{2}}^{Q^{2}} dt' \Gamma(Q^{2},t') \Delta_{h}(t',t_{0}),
\end{equation}
\begin{equation}\label{eq:secondary}
\sigma^{\text{S}}(Q^{2},Q_{0}^{2}) = C^{\text{S}} \int_{Q_{0}^{2}}^{Q^{2}} dt \Gamma(Q^{2},t) \Delta_{h}(t,t_{0}) \int_{Q_{0}^{2}}^{t} dt' \Gamma(t,t') \Delta_{h}(t',t_{0}),
\end{equation}
\end{subequations}
where $C^{P}$, $C^{S}$ are prefactors of roughly the same order that depend on the hard process, $Q$ is the scale of the hard process, $Q_{0}$ is the scale of the daughter particle, and $\Gamma(Q^{2},t)$ is obtained by integrating the splitting functions over the energy fractions $x$ and $z$. Notice that the two equations only differ in the upper limit of the second integral. Plugging the Sudakov factors found earlier and the integrated splitting functions $\Gamma(Q^{2},t)$ in Eqs. (\ref{eq:primary}) and (\ref{eq:secondary}), one can show that for both the cubic and the quartic Higgs interactions we have
\begin{equation}\label{eq:PVsS}
\frac{\sigma^{\text{S}}(Q^{2},t_{0})}{\sigma^{\text{P}}(Q^{2},t_{0})} = \frac{C^{\text{S}}}{2C^{\text{P}}}.
\end{equation}

This implies that both primary and secondary emissions have roughly similar magnitudes. This is hardly surprising as our results seem to suggest that pure Higgs events will mostly undergo a single splitting, thus primary and secondary emissions become indistinguishable, as all emitted Higgses (including the one along the "hard" line) are soft. This high-level comparison seems to suggest that there is an $O(1)$ correction to our earlier results. On the other hand, it also seems to suggest that other splitting topologies should not be drastically different from the ones resummed through the Sudakov factor, which provides further evidence that the probability of splitting is independent of the topology of the process, and that the Higgs sector should still have good behavior at high energies. Therefore, our conclusions remain valid.

\section{Discusson, Conclusions and Outlook}\label{Chap:Conclusions}
In this paper, we tackled the issue of multi-Higgs production at high energies. It is commonly suggested in the literature that due to the factorial growth in the amplitudes of $n$-Higgs production ($\mathcal{A}_{n} \sim n!$), the number of Higgses produced at high energy should be large, leading to a breakdown in perturbation theory and violation of unitarity, thereby signaling the emergence of new physics at these energy scales. Here we approached this issue from a different angle. We developed a phenomenological approach by defining the splitting functions and the Sudakov factors for the Higgs cubic and quartic interactions. Then we generalized the method of generating functionals employed in the QCD sector to pure Higgs events, and we defined several Higgs jet observables and used them to show that the pure Higgs sector should exhibit good UV behavior. We found that on average, the number of Higgses produced at high energy should remain low. This good UV behavior is mainly a result of the weak couplings of the Higgs cubic and quartic interactions which render the probability of the Higgs splitting to other Higgses low even at high energy.

Our results are in stark contrast with the results found for multi-Higgs production at or near threshold at high energies, such as the Higgsplosion proposal. We conjecture that the breakdown of perturbation theory and the violation of unitarity one finds in such a case are probably artifacts of applying perturbation theory where it is not valid, and of assuming the double-scaling limit (which implicitly assumes a large $n$ ab initio) rather than a sign of new physics. We showed that although our treatment is approximate, as we are resumming a subset of the total $n!$ Feynman diagrams and we are working in the collinear limit, it nonetheless suggests that the Higgs sector at high energies should remain under control and well-described by the SM predictions. We argue that including other topologies would not drastically alter our conclusions as the splitting functions are independent of these topologies, and the probability of splitting remains low at high energies.

We showed that for all energy scales of interest, the Higgs cubic splitting is dominant and that the quartic one is negligible. This is due to the extra phase space suppression in the quartic case relative to the cubic one. We also showed that secondary Higgs emissions are comparable to the primary ones but do not significantly affect our results. We also studied the Higgs scaling pattern and found that pure Higgs splittings follow a Poisson pattern.

The observables developed in this paper can be helpful in studying the Higgs production at high energies, and the formalism developed in this paper can be readily applied to the rest of the EW sector. Recently, Chen \textit{et. al.} \cite{Chen:2016wkt} calculated the splitting functions for all cubic interactions in the EW sector. Thus, the generating functional method can be used to define the jet observables for the rest of the EW sector. EW jets and EW corrections will become more important as the energy scale of colliders increases, especially for the 100-TeV FCC. We intend to extend our analysis to the rest of the EW sector in future work.

\section*{Acknowledgments}
I would like to thank John Terning and Ali Shirazi for the valuable discussions. I also thank Nima Akrani-Hamed, John Conway, and Robin Erbacher for answering my questions. This work has been carried out thanks to the support of the OCEVU Labex (ANR-11-LABX-0060) and the A*MIDEX project (ANR-11-IDEX-0001-02) funded by the "Investissements d'Avenir" French government program managed by the ANR.
\appendix

\section{Derivation of the Cubic Higgs Generating Funcional}\label{App1}

Starting with Eq. (\ref{eq:GFEvolation}), and using the cubic Higgs splitting function given in Eq. (\ref{eq:3HSplitting3}), the generating functional is given by:
\begin{equation}\label{eq:GFderivation1}
\Phi(t) = \Delta(t,t_{0})\Phi(t_{0}) + \Bigg(\frac{3\sqrt{2} v \lambda}{16\pi} \Bigg)^{2} \int_{t_{0}}^{t} \frac{dt'}{t^{'2}}\Delta(t,t')\int_{0}^{1}dz z(1-z) \Phi(z^{2}t')\Phi((1-z)^{2}t').
\end{equation}

We can see from the Sudakov factor of the cubic Higgs in Eq. (\ref{eq:3HSudakov1}) that at high energy, it becomes almost constant. Thus, we can neglect the $z$-dependence of the generating functionals and pull them out of the $z$-integral. This leaves $\int_{0}^{1} dz z(1-z) = 1/6$. In addition, notice $\Delta(t,t') = \Delta(t,t_{0})/\Delta(t',t_{0})$. Thus, Eq. (\ref{eq:GFderivation1}) simplies to
\begin{equation}\label{eq:GFderivation2}
\Phi(t) \simeq \Delta(t,t_{0}) \Phi(t_{0}) + \Bigg( \frac{\sqrt{3}v\lambda}{16\pi} \Bigg)^{2} \Delta (t,t_{0}) \int_{t_{0}}^{t} \frac{dt'}{t^{'2}} \frac{\Phi^{2}(t')}{\Delta(t',t_{0})}.
\end{equation}

Differentiating both sides w.r.t. $t$ and then dividing by $\Phi(t)$, we obtain a simple differential equation for the generating functional
\begin{equation}\label{eq:GFderivation3}
\frac{d\Phi(t)}{\Phi(t)} = \frac{d\Delta(t,t_{0})}{\Delta(t,t_{0})} + \Bigg( \frac{\sqrt{3}v\lambda}{16\pi} \Bigg)^{2} \frac{\Phi(t)}{t^{2}}dt.
\end{equation}

Integrating both sides from $t_{0}$ to $t$ and noting that $\Delta(t_{0},t_{0}) =1$, we obtain the following expression for the generating functional
\begin{equation}\label{eq:GFderivation4}
\Phi(t) = \Phi(t_{0}) \Delta(t,t_{0}) \text{exp} \Bigg[ \Big( \frac{\sqrt{3} v \lambda}{16 \pi} \Big)^{2} \int_{t_{0}}^{t}\frac{dt'}{t^{'2}} \Phi(t') \Bigg].
\end{equation}

By definition, the generating functional evaluated at $t_{0}$ describes jets that have no opportunity of splitting, thus $\Phi(t_{0}) \equiv u$. Given Eq. (\ref{eq:3HSudakov1}), we can write Eq. (\ref{eq:GFderivation4}) as
\begin{equation}\label{eq:GFderivation5}
\Phi(t) = u \hspace{1mm}\text{exp} \Bigg[ \Big(\frac{\sqrt{3}v\lambda}{16\pi}\Big)^{2} \int_{t_{0}}^{t} \frac{dt'}{t^{'2}} \Big( \Phi(t')-1\Big) \Bigg].
\end{equation}

Since $\int_{t_{0}}^{t} \frac{dt'}{t^{'2}}$ is dominated near $t' \sim t_{0}$, we can approximate $\Phi(t') \approx \Phi(t_{0}) = u$. Thus, we can further simpify Eq. (\ref{eq:GFderivation5})
\begin{align}
\Phi(t) & =u \hspace{1mm} \text{exp} \Bigg[ -(1-u)\Big(\frac{\sqrt{3}v\lambda}{16\pi}\Big)^{2} \int_{t_{0}}^{t} \frac{dt'}{t^{'2}}  \Bigg] \nonumber \\
& = u \hspace{1mm} \Bigg\{\text{exp} \Bigg[ -\Big(\frac{\sqrt{3}v\lambda}{16\pi}\Big)^{2} \int_{t_{0}}^{t} \frac{dt'}{t^{'2}}  \Bigg] \Bigg\}^{1-u} \nonumber \\
& = u \hspace{1mm} \Big[ \Delta(t,t_{0}) \Big]^{1-u}.
\end{align}

\end{document}